\documentclass[10pt]{article}
\usepackage{lineno}
\usepackage[letterpaper, margin=1in]{geometry}
\usepackage{graphicx}
\usepackage{subcaption}
\usepackage[range-phrase={--},per-mode = single-symbol]{siunitx}
\usepackage{chemformula}
\usepackage[version=4]{mhchem}
\usepackage{hyperref}
\usepackage{cleveref}
\usepackage{authblk}
\usepackage{varwidth}
\usepackage{xcolor}
\usepackage[utf8]{inputenc}
\usepackage[T1]{fontenc}
\usepackage[normalem]{ulem}
\usepackage{adjustbox}
\usepackage{amsmath}
\usepackage{lmodern} 
\usepackage{hyphenat}
\usepackage[
  style=numeric-comp,
  backend=biber,
  sorting=none,
  giveninits=true,
  maxnames=3,
  doi=true,
  url=false,
  isbn=false,
  eprint=true
]{biblatex}

\AtEveryBibitem{%
  \clearlist{language}
  \clearfield{issn}%
  \clearfield{month}
  \clearfield{urldate}
}

\AtEveryBibitem{%
  \iffieldundef{doi}{}{\clearfield{url}}%
}

\DeclareFieldFormat{doi}{%
  \mkbibacro{DOI}\addcolon\space\href{https://doi.org/#1}{\nolinkurl{#1}}%
}
\DeclareFieldFormat{eprint:arxiv}{%
  arXiv\addcolon\space\href{https://arxiv.org/abs/#1}{#1}%
}

\begin{document}

\title{Better without U: Impact of Selective Hubbard U Correction on Foundational MLIPs}

\author{Thomas Warford$^1$, Fabian L.~Thiemann$^2$ and Gábor Csányi$^{1,3,*}$}

\affil{$^1$Engineering Laboratory, University of Cambridge, Trumpington Street, Cambridge, CB2 1PZ, UK}

\affil{$^2$Microsoft Research AI for Science,  Cambridge, CB1 2FB, UK}

\affil{$^3$Max Planck Institute for Polymer Research, Ackermannweg 10, Mainz, Germany}

\affil{$^*$Corresponding author: \href{mailto:gc121@cam.ac.uk}{gc121@cam.ac.uk}}

\maketitle

\begin{abstract}
The training of foundational machine learning interatomic potentials (fMLIPs) relies on diverse databases with energies and forces calculated using ab initio methods. We show that fMLIPs trained on large datasets such as MPtrj, Alexandria, and OMat24 encode inconsistencies from the Materials Project's selective use of the Hubbard U correction, which is applied to certain transition metals only if O or F atoms are present in the simulation cell. This inconsistent use of +$U$ creates two incompatible potential-energy surfaces (PES): a lower-energy GGA surface and a higher-energy GGA+$U$ one. When trained on both, MLIPs interpolate between them, leading to systematic underbinding, or even spurious repulsion, between $U$-corrected metals and oxygen- or fluorine-containing species. Models such as MACE-OMAT and -MPA exhibit repulsion between $U$-corrected metals and their oxides, limiting their value for studying catalysis and oxidation. We link the severity of this pathology to the oxygen number density in $U$-corrected training configurations. This explains why OMAT-trained models are most affected and suggests the issue might worsen as expanding future datasets increasingly include configurations with low oxygen content, such as those generated through combinatorial exploration of multi-element or defect-containing systems.

Our simple per-$U$-corrected-atom shift aligns PBE+$U$ and PBE energies for identical structures, yielding a smoother PES compared to existing correction schemes, which target phase diagram accuracy. As a result, models trained on datasets with our shift applied exhibit smaller mean absolute errors for the adsorption energies of oxygen on $U$-corrected elemental slabs. Since datasets omitting +$U$ entirely (e.g. MatPES, MP-ALOE) avoid these pathologies, we recommend excluding +$U$ in future fMLIP datasets. For existing datasets, our post-hoc correction provides a low-cost improvement.
\end{abstract}

\section{Introduction}
Machine learning interatomic potentials (MLIPs) enable large-scale simulations with near first-principles accuracy~\cite{behler_generalized_2007, bartok_gaussian_2010,rupp_fast_2012, thompson_spectral_2015, musil_machine_2018, zhang_deep_2018,  zuo_performance_2020, stocker_machine_2020,  deringer_origins_2021,unke_machine_2021,kapil_first-principles_2022, 
jaffrelot_inizan_scalable_2023, witt_acepotentialsjl_2023,     zeng_mechanistic_2023,   ko_recent_2023} 
enabling, for example, the device-scale atomistic simulation of the entire write/erase cycle in a phase-change memory device~\cite{zhou_full-cycle_2025}. Recently, foundational MLIPs (fMLIPs) trained on large, chemically diverse datasets have emerged~\cite{
deng_chgnet_2023, batatia_foundation_2024, fu_learning_2025}. While many models are trained on limited chemical domains, fMLIPs aim to represent a single potential energy surface (PES) spanning many chemical systems, making them broadly applicable across materials classes. Their continued development requires careful consideration of the underlying quantum-mechanical reference data.

Since 2011, the Materials Project (MP)~\cite{horton_accelerated_2025} has provided standardized density functional theory (DFT) settings. These settings were calibrated long before the emergence of the first fMLIPs, with the goal of accurately predicting phase diagrams and material properties rather than producing a smooth PES. Regardless, the MP standards, which use the PBE~\cite{perdew_generalized_1996} exchange-correlation functional to balance accuracy and computational cost, have since been widely adopted by fMLIP datasets: MPtrj (constructed directly from MP relaxations), Alexandria, and OMat24 all use these settings, enabling MLIPs to be trained across them simultaneously~\cite{deng_chgnet_2023,schmidt_improving_2024,barroso-luque_open_2024}. Indeed, almost all of the most widely-used MLIP architectures, including  MACE~\cite{batatia_mace_2023}, eSEN~\cite{fu_learning_2025}, EquFlash~\cite{lee_flashtp}, Nequip~\cite{batzner_e3-equivariant_2022}, ORB~\cite{neumann_orb_2024}, Allegro~\cite{musaelian_learning_2022}, GRACE~\cite{bochkarev_graph_2024} and DPA~\cite{zhang_dpa-1_2022} are trained on such data.

The mismatch between MP's design and MLIPs' needs manifests most severely in the treatment of correlated electrons. Like all GGAs, PBE suffers from significant self-interaction error, particularly for localized states such as $d$ orbitals in transition-metal oxides and fluorides. This can be mitigated by applying a Hubbard $U$ correction~\cite{anisimov_band_1991,anisimov_first-principles_1997}. In the GGA+$U$ approach, as formulated by Dudarev \textit{et al.}~\cite{dudarev_electron-energy-loss_1998}, \emph{non-negative} terms are added to the total energy, $E_{\text{DFT}}$, for each $U$-corrected site
\begin{equation}
        E_{\text{DFT+U}} = E_{\text{DFT}} + \sum_{I,\sigma} \frac{U^I}{2} \text{Tr}\left[\mathbf{n}^{I\sigma} ( \mathbf{1} - \mathbf{n}^{I\sigma} )\right]
    \end{equation}
where $U^I$ is the Hubbard parameter for site $I$ and $\mathbf{n}^{I\sigma}$ is the on-site occupation matrix, defined as
    \begin{equation}
        n^{I\sigma}_{m,m'} = \sum_{i} f^\sigma_{i} \langle \psi^{\sigma}_{i} | \phi^I_{m'} \rangle \langle \phi^I_m | \psi^{\sigma}_{i} \rangle
    \end{equation}
    where $\psi^{\sigma}_{i}$ denotes a Kohn-Sham orbital with spin $\sigma$ and $\phi^I_m$ denotes localized atomic orbitals with angular momentum quantum numbers $l$ and $m$. $f^\sigma_i$ denotes the occupation of said Kohn-Sham orbital. Since the eigenvalues $\lambda^{I\sigma}_m$ of $\mathbf{n}^{I\sigma}$ lie in $[0,1]$, each term $\lambda^{I\sigma}_m(1-\lambda^{I\sigma}_m)$ is non-negative and vanishes only for integer occupations. The $+U$ correction therefore penalizes fractional occupancies and energetically favors localized, integer-occupied $l$ orbitals~\cite{leiria_campo_jr_extended_2010}. GGA+$U$, with fitted $U$ values, improves the description of correlated oxides and fluorides, reducing redox energy errors which can exceed 1 eV with plain GGAs~\cite{zhou_first-principles_2004, wang_oxidation_2006}. However, it performs poorly for metals, where the electrons are delocalized~\cite{jain_formation_2011, wang_oxidation_2006}. Since no single $U$ setting works well across both regimes, the Materials Project applies the correction only to $d$ orbitals of selected transition metals (V, W, Fe, Ni, Co, Cr, Mo, Mn) when oxygen or fluorine is present~\cite{materialsproject_hubbardu}.

\begin{figure}
        \centering
        \includegraphics[width=0.95\linewidth]{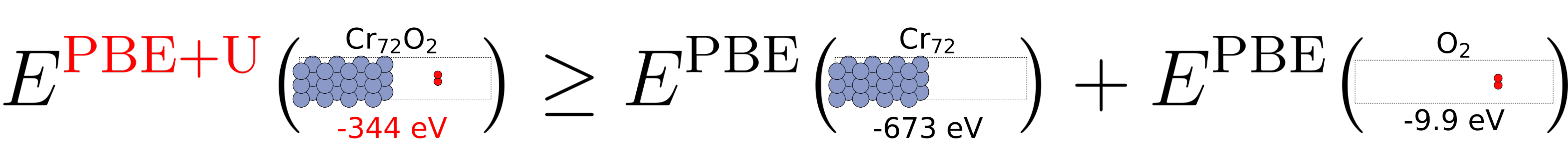}
        \caption{The high energy of a Materials Project (\texttt{MPRelaxSet}) DFT calculation for a chromium surface with a distant O$_2$ molecule arises because the +$U$ correction is applied only in this combined system, not in the separate surface or O$_2$ calculations. The PBE energy for the combined system is -683 eV.}
        \label{fig:inequality}
    \end{figure}

This selective use of $+U$ creates two distinct PES: a lower-energy GGA surface for systems without oxygen or fluorine, and a higher-energy GGA+$U$ surface when these elements are present. As shown in Figure~\ref{fig:inequality}, the sum of the energies of an O$_2$ molecule and a $U$-corrected metal slab in their own cells is always lower than the energy of the combined system, since the $U$ correction is applied only in the latter case. An MLIP with a finite cutoff radius must interpolate between these surfaces, leading to overly positive energy changes when O or F atoms approach a ‘pure’ $U$-corrected metal. To date, fMLIP development has largely overlooked these inconsistencies, with all models except CHGNet~\cite{deng_chgnet_2023} trained on the raw energies of mixed PBE/PBE+$U$ datasets. Less empirical approaches exist for determining system-specific $U$ values~\cite{cococcioni_linear_2005, kulik_density_2006}, but these do not fix the issue of energies at different $U$ being incompatible. 

Two strategies have emerged to avoid or mitigate the impact of these inconsistencies, one being to omit +$U$ entirely, as done in the MatPES and MP-ALOE datasets~\cite{kaplan_foundational_2025,kuner_mp-aloe_2025}. These studies show that models trained on the smoother PES that is obtained yield more stable molecular dynamics than those trained on MPtrj. 
Alternatively, one might adjust GGA+$U$ energies to make datasets more internally consistent. Such \emph{GGA/GGA+$U$ mixing schemes} exist, but they were not developed with MLIPs in mind. Instead of trying to map GGA+$U$ energies to GGA energies, their motivation was purely thermochemical: neither GGA nor GGA+$U$ alone accurately reproduce experimental phase diagrams; GGA tends to over-stabilize metals, while GGA+$U$ over-stabilizes oxides. To exploit the complementary strengths of each method, Jain \textit{et al.}~\cite{jain_formation_2011} proposed a scheme that shifts GGA+$U$ energies to make energy differences between calculation types meaningful. Specifically, constant per-element corrections are added for each $U$-corrected site in a cell, and these shifts are fitted to minimize the mean-squared error between experimental and computed oxide formation enthalpies. During fitting, a constant shift is also applied to O$_2$ molecules to correct for their overbinding in GGA~\cite{wang_oxidation_2006, patton_simplified_1997}, another consequence of self-interaction error. Using shifted GGA+$U$ energies in combination with GGA energies resulted in reduced formation enthalpy errors and a larger fraction of experimentally stable phases being correctly identified compared to using GGA or GGA+$U$ alone. 

The Wang \textit{et al.} scheme, adopted by the Materials Project for phase-stability analyses, extends this approach~\cite{wang_framework_2021}. In addition to per-$U$-corrected-site energy shifts, it introduces fitted per-anion corrections for oxygen and other elements when their oxidation state is estimated to be negative, rather than using a fixed correction for O$_2$.
Crucially, neither scheme was designed to yield a smooth PES; they were fitted to better reproduce experimental phase diagrams. The anion corrections of Wang \textit{et al.} are a clear reflection of this design choice: the correction is applied in a binary way, introducing discontinuities in the PES that are problematic for MLIPs.

Despite the widespread adoption of these mixed PBE/PBE+$U$ datasets for MLIP training, their suitability for this purpose remains unexplored. Millions of compute hours have been spent creating and training on mixed PBE/PBE+$U$ datasets, yet no study has systematically investigated the impact of training directly on these raw energies or the extent to which existing correction schemes mitigate the resulting problems.

In this work, we show that the Materials Project’s Hubbard $U$ scheme leads to severe pathologies when MLIPs are trained directly on raw energies. By computing DFT adsorption energies of oxygen on elemental slabs, we show that $U$-affected models predict spurious underbinding between $U$-corrected metals and oxygen or fluorine. Additionally, we show that these models systematically underestimate $U$-corrected metal-oxide adhesion energies. Most strikingly, we observe unphysical repulsion emerges in both adsorption and interface cases for the MACE-OMAT and MACE-MPA models. Despite all datasets employing the Materials Project's Hubbard $U$ parameters, these models suffer more severe artifacts than MACE-MP: models trained on the much larger OMAT dataset are most affected, while MPtrj-trained models are least. We trace these differences to variations in the oxygen number density of GGA+$U$ configurations.

We show models trained on MatPES (which omits +$U$) do not have these pathologies, whilst existing mixing schemes reduce their severity to a limited extent. We also introduce our own mixing scheme, similar to that of Jain \textit{et al.}, where per-$U$-atom shifts are fitted to minimize the mean-squared difference between shifted PBE+$U$ and PBE energies. We demonstrate that training on data with our scheme applied results in a significantly lower MAE for the binding energies of oxygen on $U$-corrected elemental slabs versus the Wang \textit{et al.} scheme. As a consequence of our investigation, we recommend that future fMLIP datasets avoid Hubbard U, while our post-hoc correction provides a low-cost fix for existing datasets.

\section{Numerical Experiments}\label{sec:experiments}
    We now perform controlled experiments to characterize the pathologies arising from the mixing of PBE and PBE+$U$ data. 
    In this context, we refer to models trained directly on raw PBE/PBE+$U$ energies as \emph{$U$-afflicted}. These models are impacted by selective application of the Hubbard U correction in the underlying data. The problems are most severe when O or F atoms enter the local environments of $U$-corrected transition metals that were previously free from these elements.

    For comparison, we include models trained on MatPES (which omits $+U$ entirely) and models trained on MPtrj with mixing schemes applied (Wang \textit{et al.} or our own). This enables direct comparison between models trained on PBE-only data, raw PBE/PBE+$U$ data, and corrected PBE/PBE+$U$ data.
    
    Our experiments progress from simple to more complex cases. We begin with oxygen adsorption on clean elemental slabs. Here the pathology is most visible: $U$-afflicted models consistently predict spurious underbinding on $U$-corrected slabs. Models trained on OMAT or Alexandria even show unphysical repulsion. This test case reveals a striking dataset dependence: OMAT-trained models perform far worse than MPtrj-trained models, despite OMAT being over 60$\times$ larger than MPtrj and both datasets using identical Hubbard $U$ settings.
    
    We then examine metal-oxide adhesion energies, which are directly relevant to corrosion and catalysis~\cite{peden_metalmetal-oxide_1991}. We show that PES discontinuities bias adhesion energy predictions, demonstrating tangible consequences of $+U$ inconsistencies.
    Finally, we examine fluorine adsorption on the Fe side of Fe-FeO slabs. By varying the thickness of Fe and FeO layers, we span surfaces from pure Fe to pure FeO. Adsorption on Fe resembles the oxygen case and reveals strong $U$-related pathologies. In contrast, adsorption on FeO is largely unaffected because models are already mimicking a GGA+$U$ PES.
    Together, these experiments reveal how +$U$ inconsistencies affect MLIPs across different scenarios and provide a basis for evaluating correction strategies.
\subsection{Models and Datasets}

Throughout our experiments, we compare a range of models to probe how dataset composition and correction schemes influence performance. We include three versions of \textbf{MACE-MP-0b3}, all trained on the configurations of \textbf{MPtrj} but with differing energy labels.
\begin{itemize}
    \item \textbf{No shift}: the original MACE-MP-0b3 trained directly on raw PBE/PBE+$U$ energies.
    \item \textbf{Wang \textit{et al.} shift}: an otherwise identical model trained on energies with the Wang \textit{et al.} scheme (implemented as \texttt{MP2020Correction} in \texttt{pymatgen}) applied. These energies are available as \texttt{\allowdisplaybreaks{corrected\_total\_energy}} in MPtrj. This model is included because the correction scheme is heavily used by the Materials Project for phase diagrams, and \textbf{CHGNet} is trained on these energies.
    \item \textbf{Our shift}: a model trained on energies shifted according to our scheme, designed to recover the smoothest possible PES. Our constant per-$U$-element energy shifts are fitted to minimize the mean squared difference between shifted PBE+$U$ energies and PBE energies for identical structures found in both MP and MatPES-PBE.
\end{itemize}

We also trained a model on MPtrj with a refitted version of the Jain \textit{et al.} shift applied. The per-$U$-corrected-element shifts and model behavior were nearly identical to those of the Wang \textit{et al.} scheme (see SM).

Additionally, we include four more models with the same architecture trained on different datasets:
\begin{itemize}
    \item \textbf{MACE-MPA-0}: trained on MPtrj combined with \textbf{salex} (a filtered version of the \textbf{Alexandria}~\cite{schmidt_improving_2024} dataset).
    \item \textbf{MACE-OMAT-0}: trained on \textbf{OMat24}~\cite{barroso-luque_open_2024}.
    \item Two \textbf{MACE-MATPES} models: obtained by fine-tuning MACE-OMAT-0 on MatPES-PBE and MatPES-r2SCAN~\cite{kaplan_foundational_2025} respectively.
\end{itemize}

In addition to these models, which all share the architecture of MACE-MP-0b3, we evaluated three models with different architectures:
\begin{itemize}
\item \textbf{MP-0-large}: a more expressive MACE model, with $L=2$ equivariant features in the output of the first layer rather than $L=1$ and a cutoff radius of 4.5 Å rather than 6 Å.
\item \textbf{CHGNet}: trained on MPtrj energies with the Wang \textit{et al.} shifts applied.~\cite{deng_chgnet_2023}
\item \textbf{eSEN-OAM-20M}~\cite{fu_learning_2025}: a state-of-the-art model on Matbench Discovery~\cite{riebesell_framework_2025} at the time of its release, trained on OMat24, Alexandria and MPtrj combined. Included to demonstrate that $U$-related pathologies arise from data rather than model design.
\end{itemize}

\subsection{Adsorption of Oxygen on Elemental Slabs} \label{sec:periodic_table}
\begin{figure}
    \centering
        \includegraphics[width=0.95\linewidth]{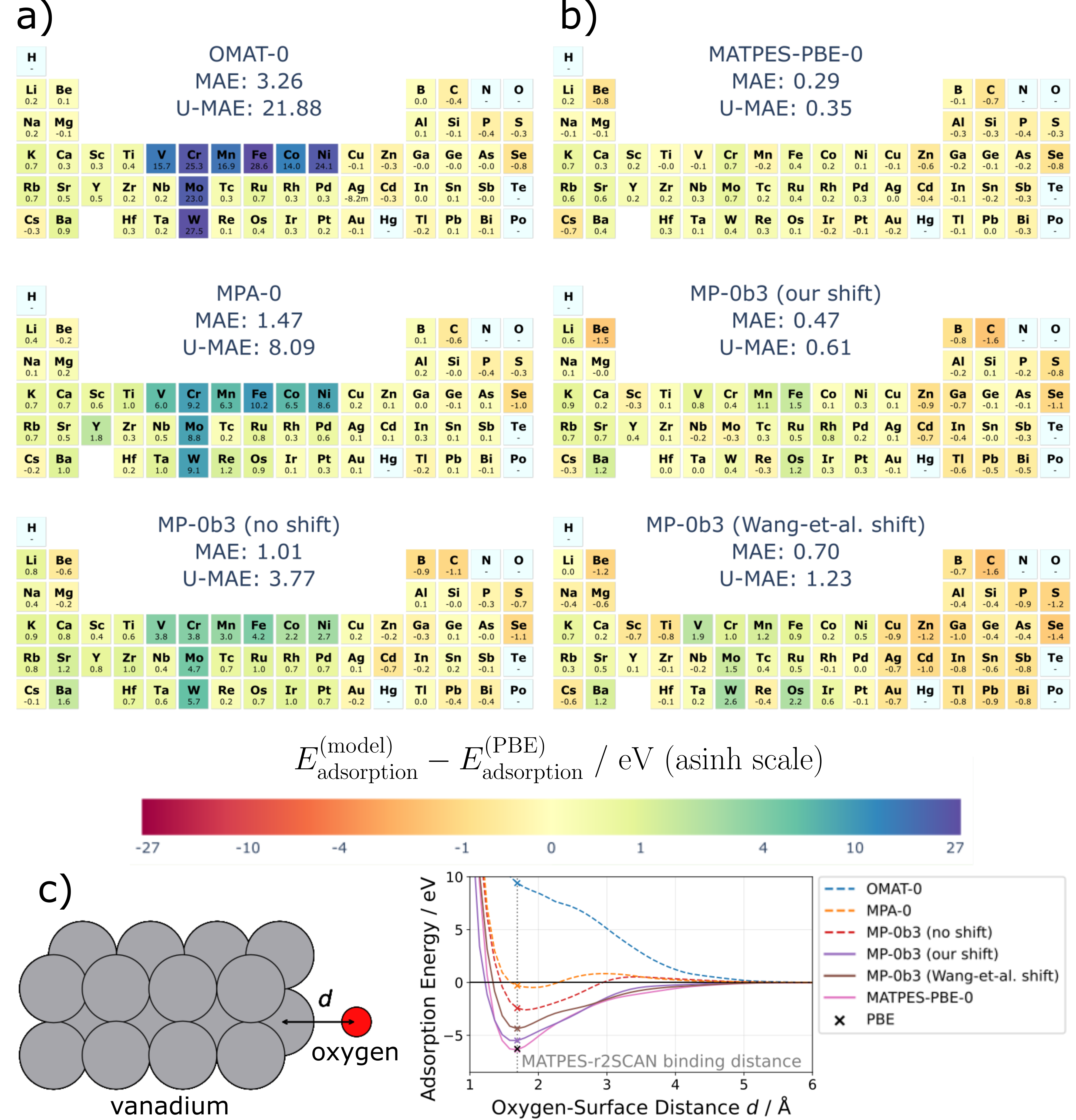}
    \caption{\textbf{Comparison of oxygen binding energies across fMLIPs}. These heatmaps show the difference between the adsorption energy predictions of various MACE models and PBE for oxygen adsorbed on top of elemental slabs. The oxygen resides on top of the site furthest along the surface's normal direction at a distance which minimizes the energy according to MACE-MATPES-r2SCAN-0, as depicted in \textbf{c)}. \textbf{a)} and \textbf{b)} show models with and without severe systematic underbinding on $U$-corrected metals respectively. MACE-OMAT-0 consistently predicts a total lack of binding whilst MACE-MPA-0 and -MP-0b3 predict barriers and significant underbinding.
    Energy differences are mapped to colours via the inverse hyperbolic sine function. Heatmaps generated with pymatviz~\cite{riebesell_pymatviz_2022}.}
    \label{fig:heatmaps}
\end{figure}    
    To demonstrate that the discussed problems are specific to $U$-affected metals and models, we performed single-point PBE calculations for oxygen placed on top of 54 elemental slabs (see Methods). The elemental slabs were obtained using the Materials Project API. Figure~\ref{fig:heatmaps} compares the adsorption energies predicted by various MACE models to those from PBE. Figure~\ref{fig:heatmaps}a shows all $U$-afflicted models exhibit severe underbinding for every $U$-corrected metal.

    The adsorption curves of different models vary markedly. Full curves and configuration plots for elements in Figure~\ref{fig:heatmaps} are provided in the ancillary files. According to MACE-OMAT-0, oxygen fails to bind to any $U$-corrected metal, with the model predicting extreme repulsion even at large distances ($>$4 Å). While MACE-MPA-0 predicts local minima for seven of the eight U-corrected metals, only the minimum shown in Figure~\ref{fig:heatmaps}c is lower in energy than the isolated slab and oxygen atom. MACE-MP-0b3 (no shift) predicts minima for all $U$-corrected metals, but with a nonphysical barrier for all except chromium. By contrast, the only non-$U$-afflicted model predicting a barrier for any $U$-corrected element is MACE-MP-0b3 (our shift), for the case of an manganese slab.
    
    The best performing model for $U$-corrected slabs is MACE-MATPES-PBE-0, since MatPES does not apply the +$U$ correction. Despite being obtained by fine-tuning MACE-OMAT-0, the model mimics one PES, rather than interpolating between a no-$U$ and +$U$ PES as the oxygen approaches. However, it still remains unclear whether pre-training on a $U$-afflicted dataset has notable impacts on fine-tuned model performance. Of the models trained on a shifted MPtrj dataset, `our shift' yields the lowest MAE on $U$-corrected surfaces by a significant margin, consistent with its objective.
\subsection{Metal-Oxide Adhesion Energies}\label{sec:interfaces}
\begin{figure}
    \centering
    \includegraphics[width=0.9\linewidth]{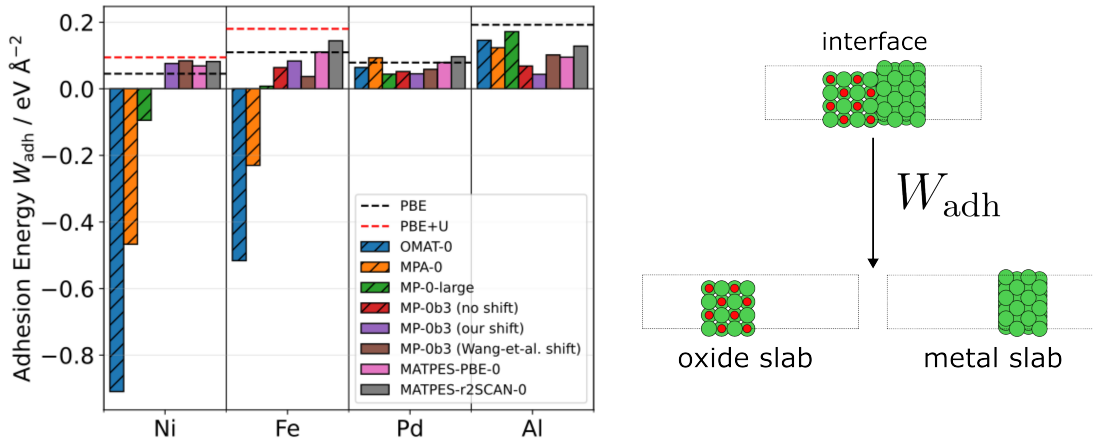}
    
    \caption{\textbf{Metal-oxide adhesion energies according to different models and DFT}. (Left) Comparison of adhesion energies for different metal-oxide interfaces. (Right) Schematic explanation of adhesion energy $W_{adh}$. Negative adhesion energies indicate unstable interfaces. Interfaces were constructed using the (100) facets of the experimentally observed face-centered cubic (FCC) metal and rocksalt oxide polymorphs.}
    \label{fig:interface_energy_scatter}
\end{figure}
    Metals typically bond readily to their oxides, forming stable interfaces. This section shows that some models trained on raw PBE and PBE+$U$ data spuriously predict unstable interfaces. This manifests in the slabs that make up Fe-FeO and Ni-NiO interfaces moving apart during relaxation with MACE-MPA-0 or MACE-OMAT-0, whilst Pd-PdO and Al-AlO interfaces are unaffected.
    
    The stability of an interface can be quantified with the adhesion energy $W_\text{adh}$
    \begin{equation}
        W_\text{adh} = \frac{1}{A} \left( E^\text{slab}_\text{metal}+ E^\text{slab}_\text{oxide} - E^\text{interface}_\text{metal-oxide}  \right)
    \end{equation}
    where $A$ is the area of the interface in the unit cell. Adhesion energy calculations typically involve relaxing both the interface and the slabs. However, since models trained on MPA and OMat24 predict repulsion between the metal and oxide, we evaluate all the models on interfaces that have been relaxed with PBE. The metal and oxide slabs are also relaxed with PBE, but with fixed lattice vectors from the relaxed interface cell so $A$ is constant. Importantly, our objective is not to reproduce experimental adhesion energies, but rather to demonstrate how $U$-afflicted models systematically fail at the atomistic simulation of these systems.
    Figure~\ref{fig:interface_energy_scatter} shows the adhesion energies according to various models and DFT. Trends from oxygen adsorption persist, with MACE-OMAT-0 being most severely affected. The MACE-OMAT-0 and -MPA-0 models predict an unstable interface for the $U$-corrected elements (Ni, Fe), whilst MACE-MP-large predicts that the Ni-NiO interface is unstable but not the Fe-FeO interface. Again, MATPES-PBE-0 is closest to the PBE reference for $U$-corrected elements.
\subsection{F Adsorption on the Fe Side of Fe-FeO Slab}\label{sec:fe-feo-f}
\begin{figure}
    \centering
    \includegraphics[width=0.8\linewidth]{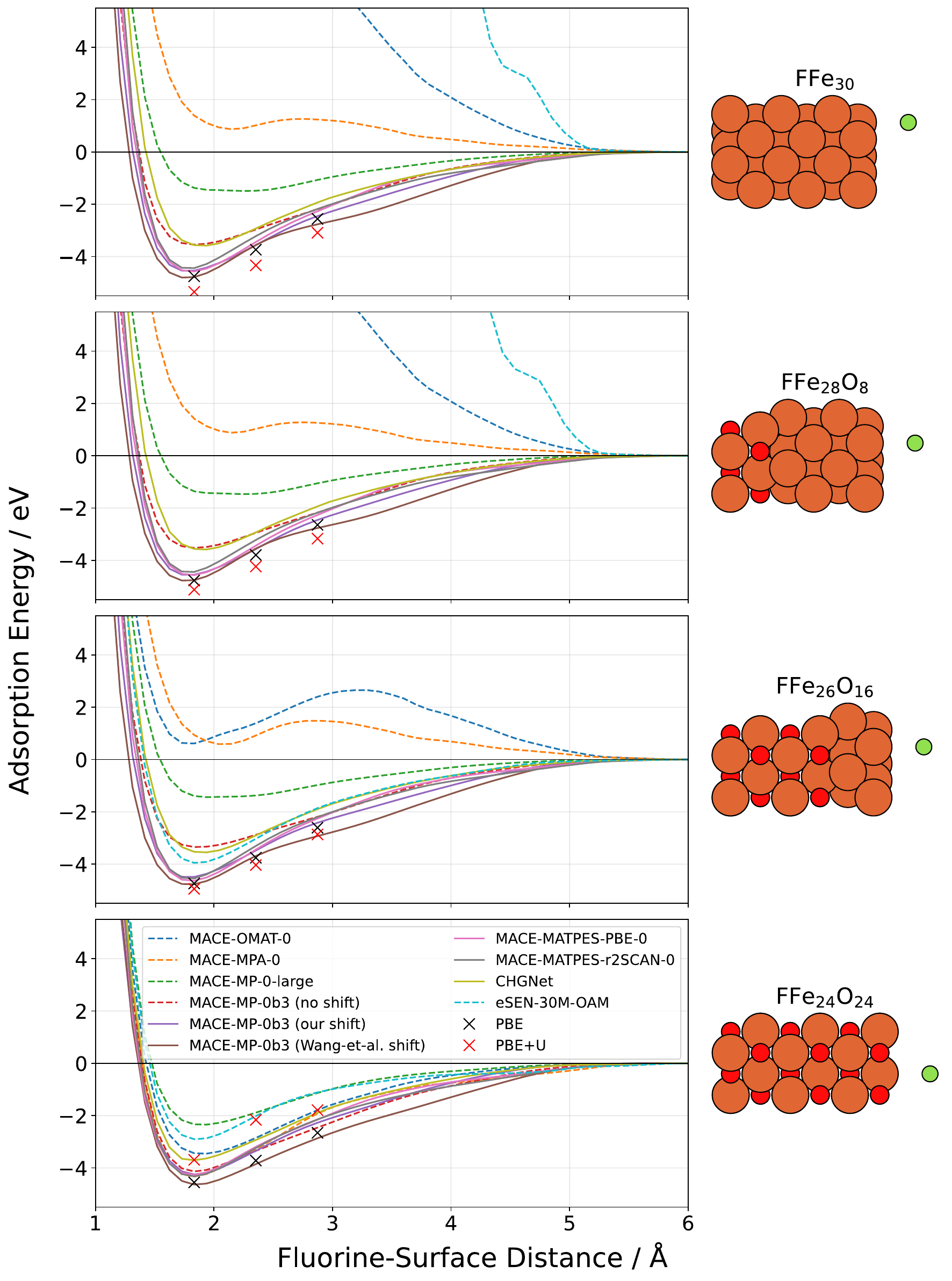}
    \caption{\textbf{Adsorption energies of fluorine on the Fe side of an Fe-FeO interface}. Although the number of (100) layers is constant, the underbinding of $U$-afflicted models is clearly worst for the all-iron slab. This occurs because iron sites with nearby oxygen atoms already produce higher-energy predictions consistent with the +$U$ regime.}\label{fig:constant_nlayers}
\end{figure}
The adsorption of fluorine onto the metal side of an Fe-FeO interface was modeled with PBE and various models. In addition to MACE models CHGNet~\cite{deng_chgnet_2023} and eSEN-30M-OAM~\cite{fu_learning_2025} are evaluated. 

Consistent with Figure~\ref{fig:heatmaps}, models trained on raw GGA/GGA+$U$ energies systematically underbind the adsorption of O and F onto the pure iron slab. For example, eSEN predicts strong repulsion as soon as the fluorine atom enters the cutoff of the iron atoms, and MACE-OMAT-0 and MACE-MPA exhibit repulsion with a smaller magnitude. These models show similar deviations for the Ni-NiO system (Figure S2).
When one of the three (100) layers is replaced with oxide, the adsorption curves remain qualitatively similar to the pure iron case. In this setup, all oxygen atoms are more than one cutoff radius (6 Å, except for MACE-MP-0-large, which has a 4.5 Å cutoff) away from the layer nearest the fluorine. Replacing two of the three (100) layers with oxide changes the results noticeably: eSEN now produces a qualitatively correct binding curve, while MACE-OMAT-0 predicts an unstable minimum. This indicates that eSEN has transitioned more abruptly to mimic the +$U$ PES, since the iron atom closest to the fluorine now has seven oxygen neighbors within 6 Å, consistent with eSEN having more layers and nonlinearities than MACE. When the slab is entirely iron oxide, all models produce the correct qualitative shape. In this case, the local environments of all iron atoms are firmly within the +$U$ regime, so the models no longer interpolate between a lower-energy (PBE) and higher-energy (PBE+$U$) potential energy surfaces as the fluorine approaches. Figures~S2-S4 show similar curves for adsorption onto Ni-NiO interfaces and no pathological behavior for Pd- and Al-oxide interfaces.
\subsection{Pathology Dependence on Dataset Structures}
\begin{figure}
    \centering
    \includegraphics[width=0.8\linewidth]{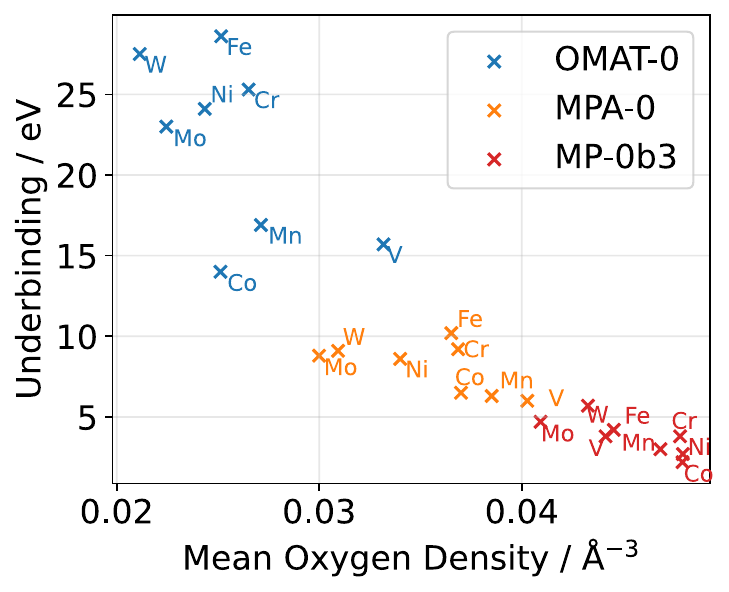}
    \caption{\textbf{Oxygen underbinding worsens with lower oxygen density in +$U$ training data.} The Y axis shows the model’s degree of oxygen underbinding relative to PBE on an elemental slab, with values identical to those in Figure~\ref{fig:heatmaps}a. Each X value corresponds to the mean oxygen number density ($N_\text{O}/V_\text{cell}$) across all configurations in the training dataset that contain both oxygen and the respective element.}
    \label{fig:o_number_density}
\end{figure}

As shown, models trained on OMat24 exhibit much more severe pathological behavior than those trained on MPtrj. Since all three $U$-afflicted datasets discussed use the Material Project's Hubbard $U$ scheme, dataset composition must play a significant role. Figure~\ref{fig:o_number_density} highlights a connection between oxygen number density and the extent of underbinding discussed in Subsection~\ref{sec:periodic_table}. As the number density of oxygen (in configurations containing the element in question) increases, the underbinding of oxygen on slabs made of said element tends to decrease. This is likely a reflection of differing distributions of local environments: $U$-corrected sites in cells with a low oxygen density are more likely to have their closest oxygen neighbors at the edges of their receptive fields. Training examples like this teach the model that whenever an oxygen comes within the cutoff radius of a $U$-corrected element, the energy must increase sharply, resulting in the extreme repulsion observed.

\section[Per Atom Energy Shift for U-Corrected Metals]{Per Atom Energy Shift for $U$-Corrected Metals}\label{sec:our_shift}
Training MLIPs on a mix of raw PBE/PBE+$U$ energies results in pathological behavior, since the MLIPs interpolate between the PBE and higher-energy PBE+$U$ potential energy surface. We now introduce a minimal correction that directly addresses the discontinuity between PBE and PBE+$U$ energy scales.

Although omitting +$U$ entirely, as in MatPES and MP-ALOE, is the most robust solution, large and widely used datasets (e.g. OMat24, Alexandria, MPtrj) already inherit the Materials Project settings. This motivates a simple, interpretable, and low-cost post-hoc correction. Our proposed per-atom energy shift addresses this need by applying constant energy shifts for each $U$-corrected metal site.

In contrast to previous GGA/GGA+$U$ mixing schemes (e.g. Jain \textit{et al.}, Wang \textit{et al.}), which were fitted to reproduce experimental formation enthalpies, our approach is designed to recover a smooth potential energy surface. This emphasis on PES continuity rather than thermochemical accuracy makes the correction better suited to training fMLIPs.

In this section we outline how the per-atom energy shifts are fitted and compare them with existing correction schemes, showing that they both confirm the origin of the $U$-induced pathologies and provide a practical improvement for large datasets based on the Materials Project settings.

Following Jain \textit{et al.}, the shifted GGA+$U$ energy is defined as
\begin{equation}
    E^\text{shifted}_{\text{GGA+}U}=E^\text{raw}_{\text{GGA+}U} + \sum_M n_M \Delta E_M
\label{eq:mixing}
\end{equation}
where $M$ denotes a metal species with a Hubbard $U$ correction, $n_M$ is the number of such atoms in a cell and $\Delta E_M$ is the fitted per-atom shift.

To fit our correction, we leveraged the relaxed structures in MATPES-PBE. Of the 108,081 ground-state structures, 25,094 have direct counterparts in Materials Project computed with PBE+$U$. This overlap enables a one-to-one comparison between PBE and PBE+$U$ energies for identical geometries. The per-atom energy shifts $\Delta E_M$ are fitted to minimize mean squared error between  $E^\text{shifted}_{\text{PBE+}U}$ and $E_\text{PBE}$.

\begin{figure}
        \centering
        \includegraphics[width=0.9\textwidth]{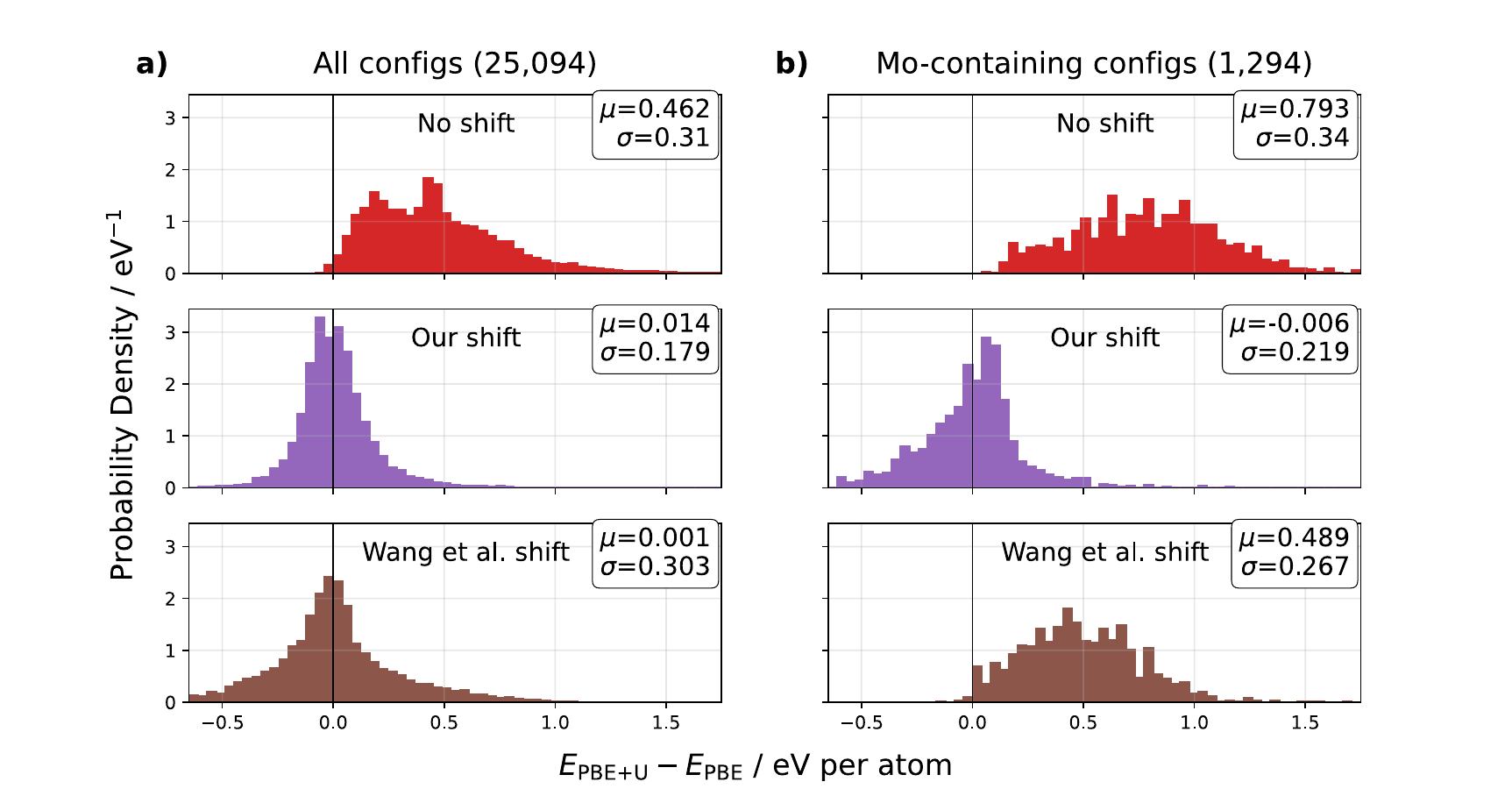}
        \caption{\textbf{Distribution of PBE+$U$--PBE energy differences.} Histograms showing the energy difference between PBE+$U$ energies from Materials Project---with no shift, our shift or the Wang \textit{et al.} shift applied---and PBE calculations from identical structures in MatPES. The Jain \textit{et al.} histograms (not shown) look near-identical to the Wang \textit{et al.} histograms, with $\mu$ and $\sigma$ both within 0.05 eV.  \textbf{a)} Distributions for  all 25,094 +$U$ structures also found in MatPES, \textbf{b)} Distributions for 1,294 +$U$ structures containing Mo.} \label{fig:hists}
    \end{figure}

Before correction, the mean difference between PBE+$U$ and PBE energies across all configurations is 0.46 eV per atom (2.5 eV per $U$-corrected atom). After applying our fitted shifts, this reduces to 0.014 eV per atom (Figure~\ref{fig:hists}), effectively reconciling the two energy scales at negligible cost.

    We next compare our shifts to those of Wang \textit{et al.}, which are optimized for thermochemical accuracy rather than PES smoothness. Consequently, while they reduce the mean offset between PBE+$U$ and PBE energies, residual inconsistencies remain that can lead MLIPs trained on such data to still exhibit a significantly distorted energy landscape.

     As shown in Figure 6, the Wang \textit{et al.} correction centers the mean PBE+$U$--PBE difference near zero, but the variance remains 1.7$\times$ larger than in our scheme. 
     Looking at configurations containing specific $U$-corrected metals explains why: elements such as V, W and Mo have large means of -0.35, 0.25 and 0.49 eV per atom respectively.
     Our per-atom shift reduces these biases to 0.010, 0.022 and -0.006 eV per atom, producing a narrower overall energy-difference distribution.

\begin{figure}
        \centering
        \includegraphics[width=0.8\linewidth]{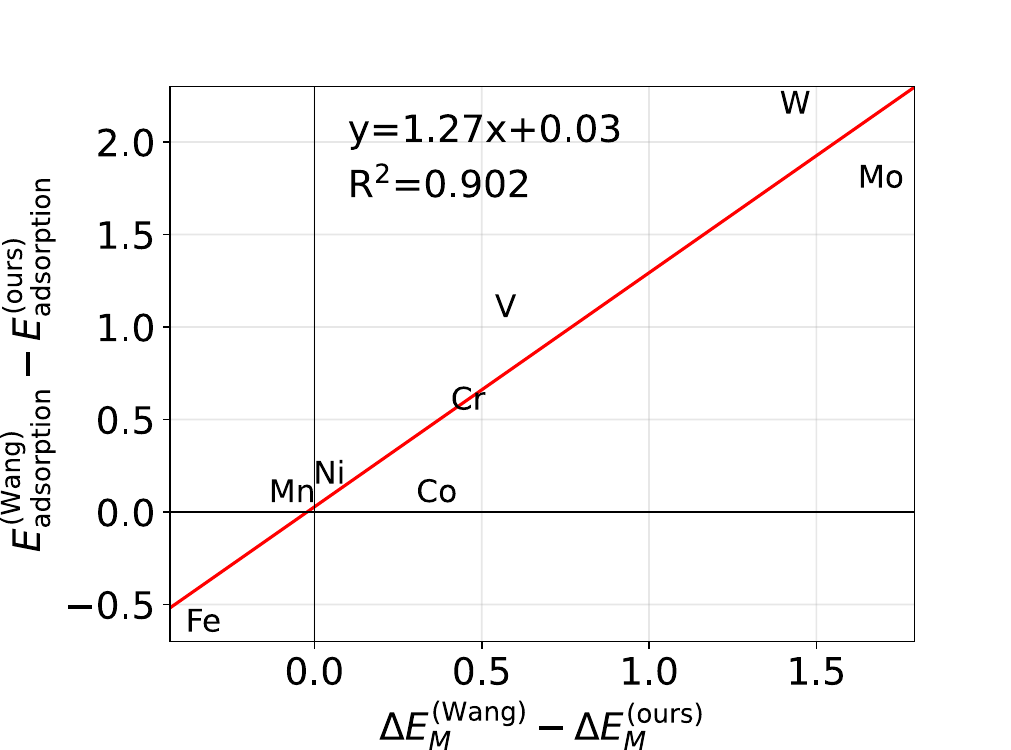}
        \caption{\textbf{Correlation between predicted adsorption energies and per-atom shifts.} The high correlation between the difference in MP-0b3 adsorption energies and the difference in per-element shifts (our correction vs. Wang \textit{et al.}) demonstrates that the improved MAE for oxygen adsorption on $U$-corrected elements (Figure~\ref{fig:heatmaps}) arises from our correction scheme targeting smoothness.}
        \label{fig:ours-wang}
    \end{figure}

    Models trained on MPtrj with our per-atom shift achieve less than half the binding-energy MAE of those trained with the Wang \textit{et al.} shift in Figure \ref{fig:heatmaps}b. The difference in binding energies correlates almost linearly with the difference in per-element shifts (Figure \ref{fig:ours-wang}), reinforcing that the energy-scale mismatch between PBE and PBE+$U$ underlies the pathological adsorption behavior.

    In summary, while existing GGA/GGA+$U$ mixing schemes remain useful for phase-diagram accuracy, our per-atom shift offers a simple and physically motivated solution to the interpolation artifacts identified in Section~\ref{sec:experiments}. By prioritizing PES smoothness, it produces more consistent energy landscapes and consequently more reliable MLIPs for molecular dynamics and other atomistic simulations.

\section{Conclusion}
We have shown that training MLIPs on datasets employing the commonly used Materials Project Hubbard $U$ scheme leads to systematic deviations in predicted energetics. Specifically, when oxygen or fluorine approach $U$-corrected metals, the predicted change in energy is consistently more positive than in PBE, resulting in the underbinding of oxygen on metal surfaces and of metal-oxide interfaces.

While we show these pathologies affect all architectures trained on affected data, their severity varies systematically across training datasets: models trained on databases with lower oxygen number density exhibit more severe effects, including spurious repulsion.
This makes addressing the issue increasingly important as materials datasets continue to expand. As larger chemical spaces are explored, whether through combinatorial sampling of multi-element compounds or by introducing oxygen- or fluorine-containing defects, the average oxygen/fluorine number density is likely to decrease further, exacerbating the problem.

For future foundational datasets and benchmarks, we suggest avoiding the selective application of +$U$, as MatPES and MP-ALOE do. For existing datasets, we have fitted and validated a per-$U$-corrected-atom energy shift that minimizes the mean-squared difference between PBE+$U$ and PBE energies for identical structures. Unlike existing GGA+$U$/GGA mixing schemes, which target thermochemical accuracy for phase diagram prediction, our scheme is designed to recover a smooth potential energy surface. In addition to supporting that the inconsistent application of the Hubbard $U$ correction is the cause of the pathological behavior, this post-hoc correction can be cheaply applied to large datasets, such as OMat24.

While avoiding the selective application of +$U$ preserves a single, continuous PES, the challenge of accurately describing systems with strongly localized electrons persists. Modern meta-GGAs like r$^2$SCAN~\cite{furness_accurate_2020} offer a significant improvement over GGAs, but still benefit from Hubbard $U$ corrections when modelling transition metal oxides~\cite{artrith_data-driven_2022, kaczkowski_comparative_2021, sai_gautam_evaluating_2018}, even if the optimal $U$ values are much smaller~\cite{long_evaluating_2020}. Ultimately, improving the universality of MLIPs will require training on higher-fidelity reference data from methods such as hybrid DFT or RPA. Given the high computational cost of these methods, multi-fidelity models~\cite{chen_learning_2020, jacobson_leveraging_2023,zhang_dpa-2_2024,shoghi_molecules_2024,kim_data-efficient_2025, zhang_graph_2025,wood_uma_2025,batatia_cross_2025}---which can leverage vast amount of lower-fidelity (GGA, metaGGA) data alongside higher-fidelity calculations---offer a promising path towards MLIPs capable of accurately modeling systems across the full breadth of chemistry and materials science.
\section{Methods}
    \subsection{Elemental Slabs}
    Adsorption energy calculations were performed for all elements in groups 1-16 and periods 1-6 with DFT-relaxed surfaces accessible via the Materials Project API~\cite{tran_surface_2016}. We queried the most stable experimentally observed elemental crystals from the Materials Project and then chose their lowest surface energy facets.
    \subsection{Adsorption Energy Calculations}
    All adsorption energies were calculated by placing the adsorbate atoms (oxygen or fluorine) on top of the surface atom lying furthest along the direction normal to the surface, at a height that minimizes the energy according to the MACE-MATPES-r2SCAN-0 model. Single point energies are calculated for the bare slab, the isolated atoms and the combined system enabling the calculation of the adsorption energy $E_\text{ads}=E_\text{combined}-E_\text{slab}-E_\text{atom}$.

    \subsection{Metal-Oxide Interfaces}
    Metal-oxide interfaces for Ni, Fe, Pd, and Al were built from the (100) facets of the experimentally observed FCC metals and their rocksalt oxides. Interface cells were generated using pymatgen’s implementation of the Zur-McGill lattice-matching scheme \cite{zur_lattice_1984}, with the maximum allowed strain set to 0.1. Since FeO and NiO exhibit antiferromagnetism in nature, among the candidate supercells, the lowest-mismatch cell that could support antiferromagnetism in the oxide was selected. All interface structures were subsequently relaxed using PBE. For adhesion energy calculations the slabs are composed of two layers of metal and two of oxide. For the adsorption of fluorine, the total number of layers is kept constant at four, whilst the number of oxide layers varies from zero to four.

    \subsection{Density Functional Theory Calculations and Relaxations}
    All density functional theory (DFT) calculations were performed in VASP 6.2.0 using the Perdew-Burke-Ernzerhof (PBE) functional~\cite{perdew_generalized_1996}.
    
    Oxygen adsorption energies on elemental slabs were calculated using the \texttt{MatPESStaticSet}, implemented in pymatgen~\cite{ong_python_2013}. A reduced cutoff energy of 520 eV was used, which is greater than the recommended minimum for all the relevant pseudopotentials.
    All other single-points and relaxations were performed using \texttt{MPRelaxSet}, the input set used to generate MPtrj, Alexandria and OMat24, with Gaussian smearing applied for compatibility with slab systems.

\section*{Acknowledgments}
This work was supported by EPSRC and IBM under Grant EP/Z531005/1. We acknowledge the Technical University of Denmark for access to the Niflheim cluster that was used for DFT calculations. We acknowledge access to the Jean Zay cluster, which was used for training MACE models, as part of the Grand Challenge: GC010815458 (Grand Challenge Jean Zay H100). TW would like to thank James Darby for helpful discussions.

\section*{Conflict of Interest}
GC is a partner in Symmetric Group LLP that licenses force fields commercially and also has equity interest in Ångström AI.

\break

\printbibliography

\end{document}


\maketitle

\tableofcontents

\break    

\section{Jain \textit{et al.} Shift}
    The Jain \textit{et al.} GGA/GGA+$U$ energy mixing scheme applies constant energy shifts for each $U$-corrected site in order to better predict phase stability. This is similar to the later approach of Wang \textit{et al.}, but the Wang scheme additionally applies \emph{anion corrections}. Anion corrections are energy shifts which are only applied to sites where the estimated oxidation state of a species is negative (or, if estimated oxidation states are unavailable, when the species is the most electronegative in the compound). Since anion corrections are applied in a binary fashion, and hence introduce energy discontinuities, they are not suited for machine learning potentials. Therefore, we also trained MACE on MPtrj with a version of the Jain \textit{et al.} shift applied to energies. 

    The original Jain corrections were fitted to calculations with different $U$ values than those used in Materials Project. To ensure consistency we refitted the energy shifts to the same dataset used by Wang et al.: 222 experimental formation enthalpies.

    Table~\ref{tab:shifts} compares the energy shifts from this work (main text), the refitted Jain scheme, and the Wang et al. scheme. The shifts applied to $U$-corrected sites are similar across the Wang and refitted Jain schemes. The main difference lies in the oxygen species. When refitting the Jain scheme a fixed shift of –0.68 eV was applied to all oxide species. This uniform shift is effectively equivalent to applying a fixed correction of +1.36 eV to the O$_2$ reference energy, as in the original Jain et al. approach, because oxide formation energies depend only on relative shifts in the O$_2$ baseline. By contrast, the Wang scheme distinguishes between oxides (–0.687 eV), peroxides (–0.465 eV), and superoxides (–0.161 eV), identified via O–O bond lengths: $<$1.35 Å (superoxide), $<$1.49 Å (peroxide), otherwise oxide.

    \begin{table}[h]
\caption{\textbf{Energy Shifts for $U$-Corrected and Anion Sites}. Anion corrections are only applied when the estimated oxidation state of a species is negative (or, if estimated oxidation states are unavailable, when the species is the most electronegative in the compound). An asterisk ($^\ast$) indicates a shift that was fixed during fitting.}
\centering
\begin{tabular}{l c c c l}
\hline
Element & This work (eV) & Jain \textit{et al.} (refitted, eV) & Wang \textit{et al.} (eV) & Applies to \\
\hline
\noalign{\smallskip}
Co & -2.004 & -1.659 & -1.638 & TM oxides and fluorides \\
Cr & -2.458 & -2.011 & -1.999 & TM oxides and fluorides \\
Fe & -1.925 & -2.333 & -2.256 & TM oxides and fluorides \\
Mn & -1.602 & -1.685 & -1.668 & TM oxides and fluorides \\
Mo & -4.895 & -3.281 & -3.202 & TM oxides and fluorides \\
Ni & -2.586 & -2.758 & -2.541 & TM oxides and fluorides \\
V  & -2.271 & -1.717 & -1.700 & TM oxides and fluorides \\
W  & -5.875 & -4.459 & -4.438 & TM oxides and fluorides \\
\hline
\noalign{\smallskip}
O           & -- & -0.68$^{*}$ & -0.687 & oxides \\
Peroxide    & -- & -0.68$^{*}$ & -0.465 & peroxides \\
Superoxide  & -- & -0.68$^{*}$ & -0.161 & superoxides \\
\hline
\noalign{\smallskip}
S  & -- & -- & -0.503 & anion cpds. \\
H  & -- & -- & -0.179 & anion cpds. \\
F  & -- & -- & -0.462 & anion cpds. \\
Cl & -- & -- & -0.614 & anion cpds. \\
Br & -- & -- & -0.534 & anion cpds. \\
I  & -- & -- & -0.379 & anion cpds. \\
N  & -- & -- & -0.361 & anion cpds. \\
Se & -- & -- & -0.472 & anion cpds. \\
Si & -- & -- &  0.071 & anion cpds. \\
Sb & -- & -- & -0.192 & anion cpds. \\
Te & -- & -- & -0.422 & anion cpds. \\
\hline
\end{tabular}
\label{tab:shifts}
\end{table}

    Models trained with the refitted Jain shift behave very similarly to those trained with the Wang shift. Since this variant does not alter any qualitative conclusions, it is not analyzed further. 
    Figure S1 shows the deviations between predicted and PBE oxygen adsorption energies across the periodic table using the Jain model. Like the Wang-shifted model, the refitted Jain model tends to underbind oxygen on $U$-corrected slabs.

    \begin{figure}
    \centering
    \includegraphics[width=0.9\linewidth]{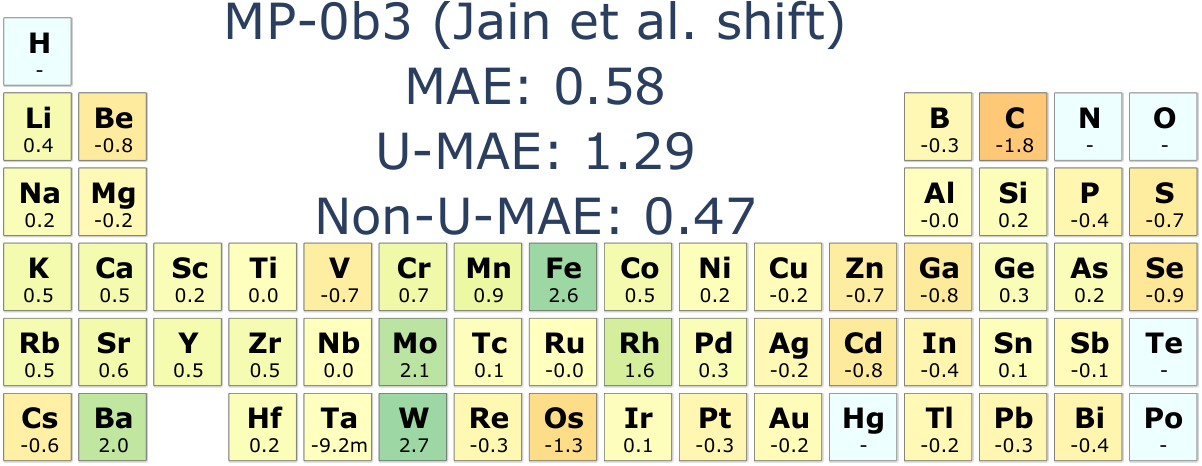}
    \caption{\textbf{Periodic table showing the difference between predicted and PBE adsorption energies in eV.} Similar to the MACE-MP-0b3 (Wang-et-al. shift) model, this model still systematically underbinds oxygen on $U$-corrected slabs.}
    \label{fig:jain}
    \end{figure}

\section{F Adsorption on the Metal Side of Metal–Oxide Slabs}
\begin{figure}
    \centering
    \includegraphics[width=0.9\linewidth]{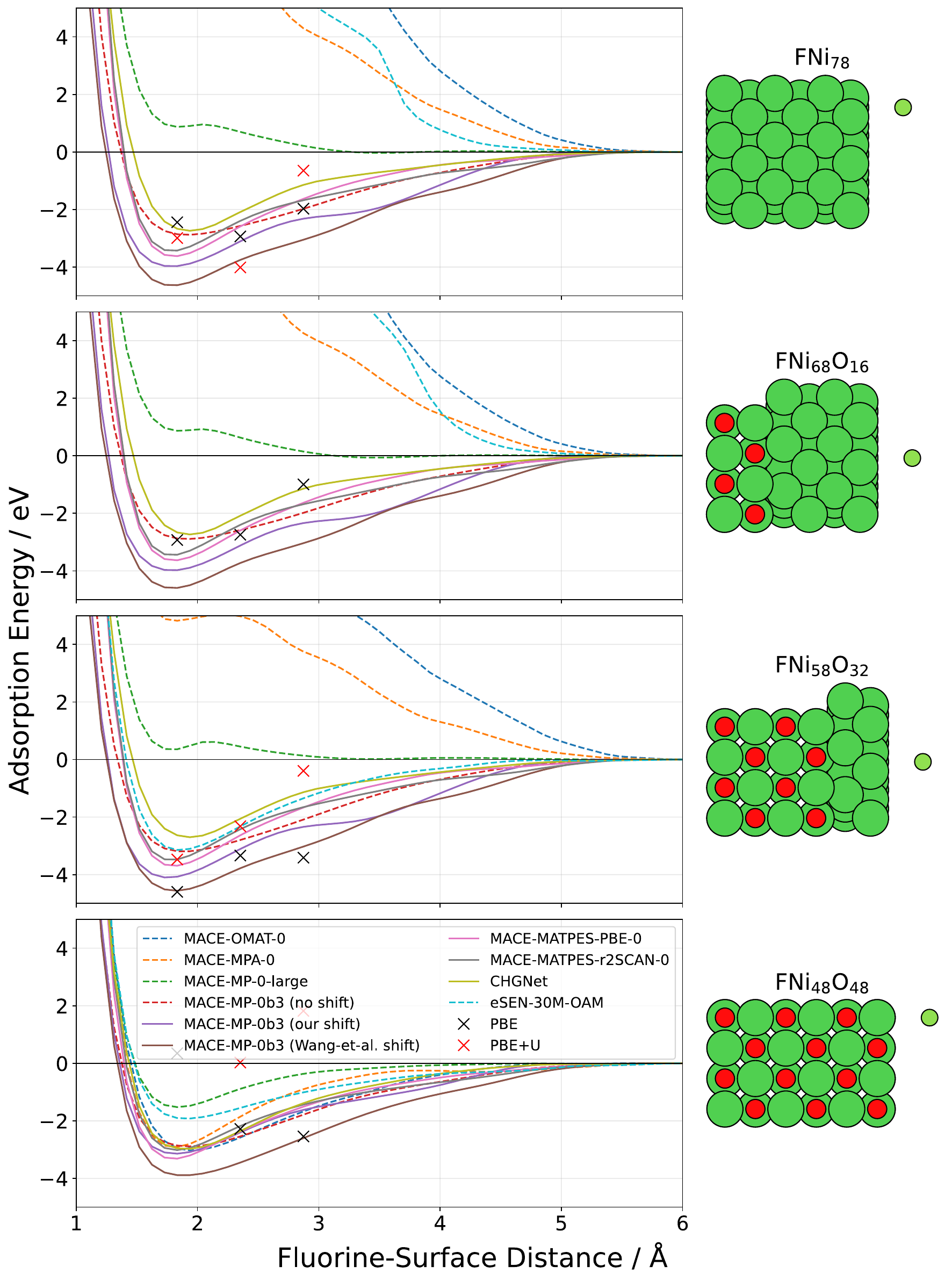}
    \caption{\textbf{Adsorption of Fluorine on the Metal Side of a Ni-NiO Slab.} The noise in DFT points is a result of calculations converging to different magnetic orderings.}
    \label{fig:Ni}
\end{figure}

\begin{figure}
    \centering
    \includegraphics[width=0.9\linewidth]{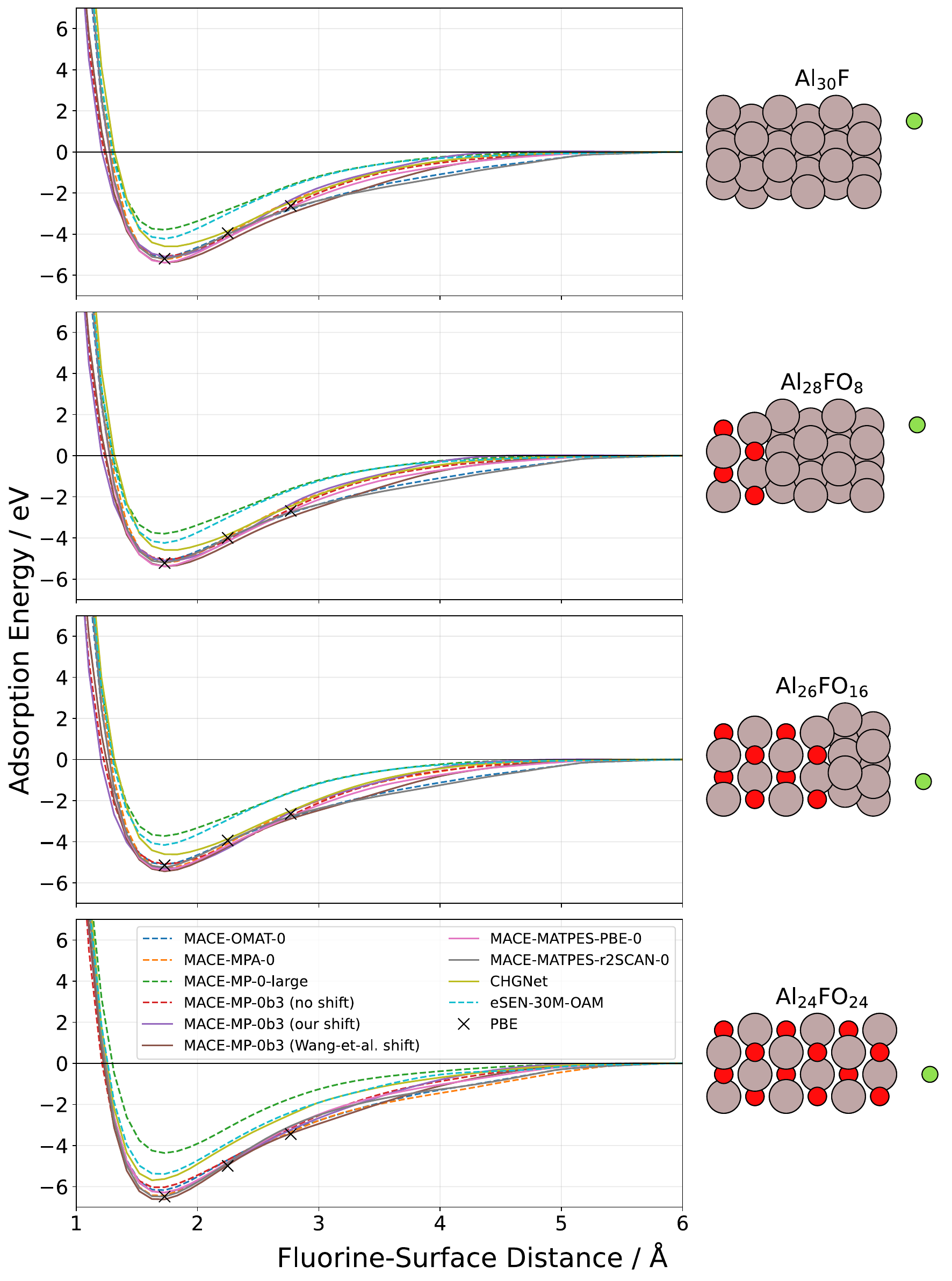}
    \caption{\textbf{Adsorption of Fluorine on the Metal Side of a Al-AlO Slab.}}
    \label{fig:Al}
\end{figure}

\begin{figure}
    \centering
    \includegraphics[width=0.9\linewidth]{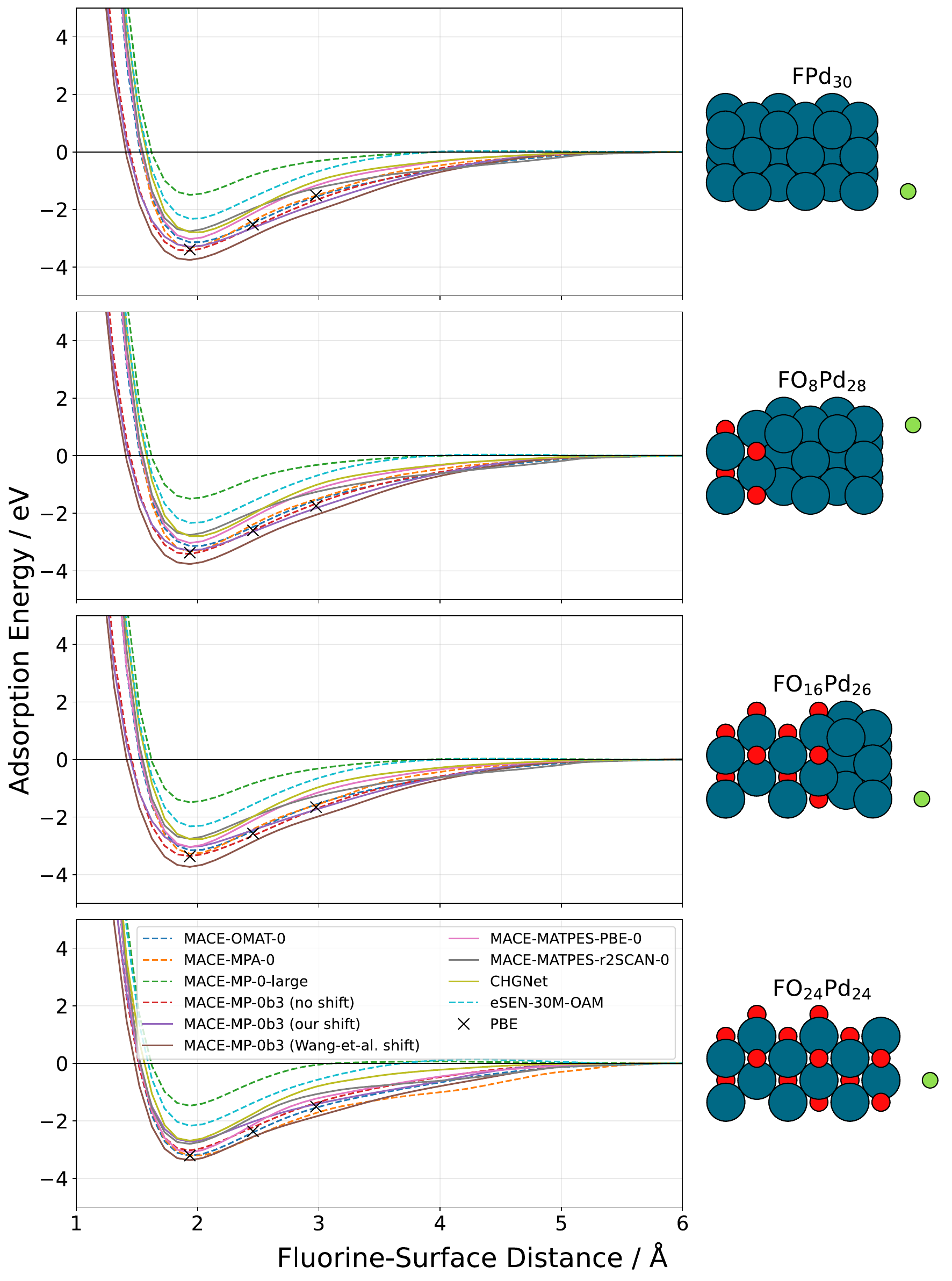}
    \caption{\textbf{Adsorption of Fluorine on the Metal Side of a Pd-PdO Slab.}}
    \label{fig:Pd}
\end{figure}

The adsorption of fluorine was also simulated on Ni-NiO, Al-AlO and Pd-PdO interfaces. Pathological behavior is observed only for $U$-afflicted models on $U$-corrected nickel (Figure~\ref{fig:Ni}). No such issues arise for aluminium or palladium. (Figures~\ref{fig:Al} and ~\ref{fig:Pd}).

For Ni-NiO, both PBE and PBE+$U$ energies show substantial noise. This originates from the sensitivity of the slab’s magnetic ordering to the fluorine position. The Materials Project’s \texttt{MPRelaxSet} initializes all magnetic moments in a ferromagnetic configuration, and depending on where the fluorine is placed, the relaxation may converge to different spin states. This inconsistency in magnetic ordering leads to variations in the reference DFT energies.

